\documentclass[jnr,crcready,onecolumn]{iosart2x}
\makeatletter
\let\numberlines@hook\relax
\makeatother

\pubyear{0000}
\volume{0}
\firstpage{1}
\lastpage{1}

\begin{document}

\begin{frontmatter}

\title{Nuclear data development at the European Spallation Source}
\runtitle{Nuclear data development at the European Spallation Source}


\author[A]{\inits{J.I}\fnms{Jose Ignacio} \snm{Marquez Damian}\ead[label=e1]{marquezj@ess.eu}%
\thanks{Corresponding author. \printead{e1}.}},
\author[A]{\inits{D.}\fnms{Douglas D.} \snm{DiJulio}},
and
\author[A]{\inits{G.}\fnms{Günter} \snm{Muhrer}},
\address[A]{Spallation Physics Group, \orgname{European Spallation Source ERIC},
Lund, \cny{Sweden}}

\begin{abstract}
Transport calculations for neutronic design require accurate nuclear data and validated
computational tools. In the Spallation Physics Group, at the European Spallation Source, we
perform shielding and neutron beam calculations to help the deployment of the instrument suite
for the current high brilliance (top) moderator, as well for the design of the high intensity bottom
moderator, currently under study for the facility.
This work includes providing the best available nuclear data in addition to improving models
and tools when necessary. In this paper we present the status of these activities, which
include a set of thermal scattering kernels for moderator, reflector, and structural materials, the development of new kernels for beryllium considering crystallite size effects,
nanodiamonds, liquid hydrogen and deuterium based on path integral molecular dynamics, and the use of the software package NCrystal to assist the development of nuclear data
in the framework of the new HighNESS project.
\end{abstract}


\end{frontmatter}


\section{Introduction}\label{s1}

Neutronic calculations for the simulation and development of moderators and reflectors at spallation neutron sources, such as the European Spallation Source (ESS) \cite{Garoby_2017}, currently under construction in Lund, Sweden, require accurate models for the interaction of thermal and cold neutrons with matter in order to get the best possible estimates for the performance of the facility. For this reason, we have launched several activities at the ESS to produce updated and revised libraries for Monte-Carlo simulations. These include improvements to the nuclear data library preparation process, implementation of small-angle neutron scattering and crystalline effects into new scattering libraries, preparation of libraries using state-of-the art molecular modelling techniques, generation of new libraries for future moderator development at the ESS, and also improvements to previous evaluations, for example in light water. In the following, we give a brief overview of these activities and present some relevant results.

\section{Nuclear data library preparation}\label{s2}

\begin{figure}[t]
\resizebox{0.6\textwidth}{!}{\includegraphics{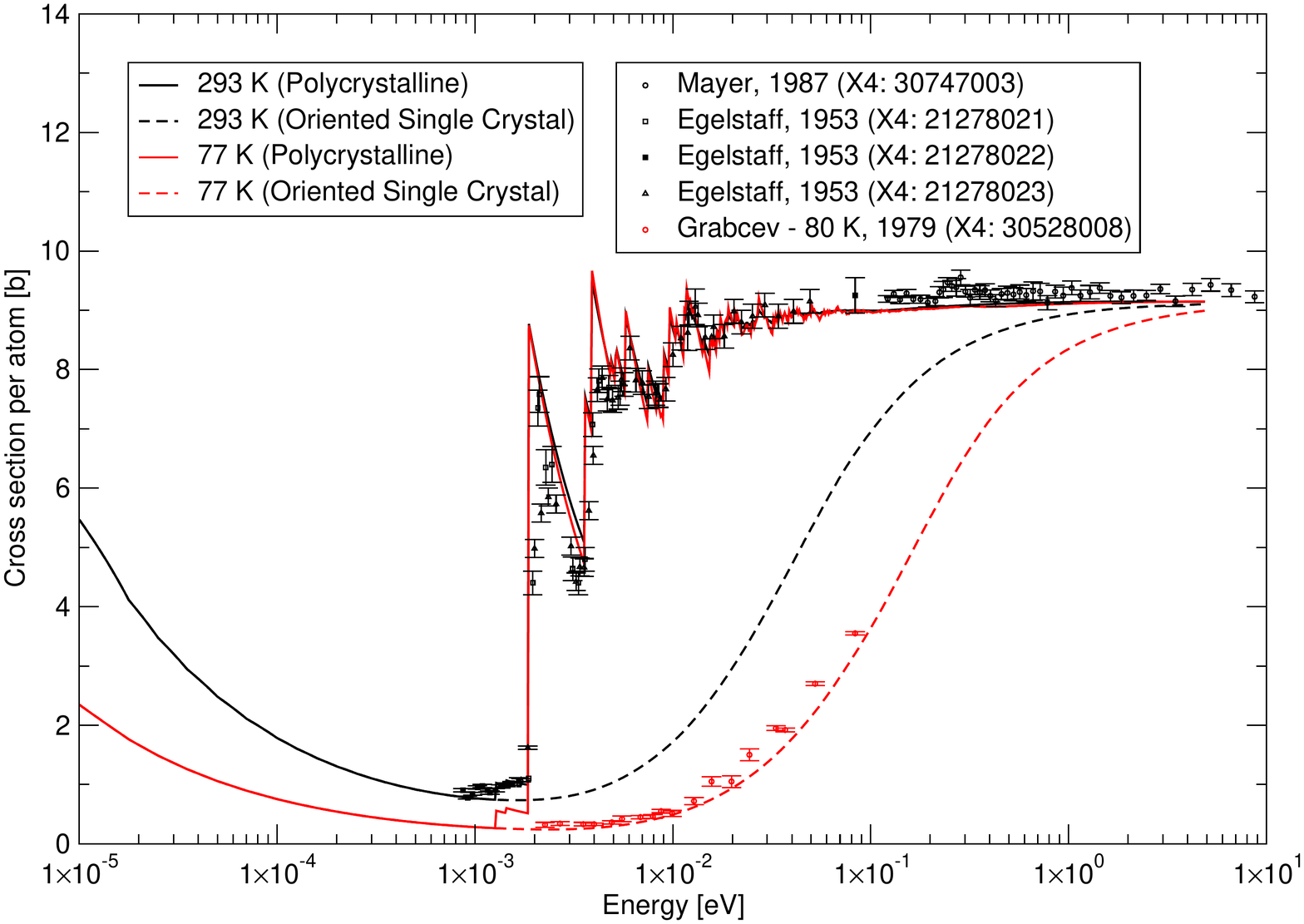}}
\caption{Total cross section for polycrystalline and oriented single crystal bismuth, compared with experimental data from EXFOR \cite{exfor}.} \label{xs_bi}
\end{figure}

In order to reduce systematic errors caused by nuclear data, we prepared a working library of thermal scattering kernels using the best available models and formats. This library contains scattering kernels for 16 materials (Table \ref{t1}) compiled from different
sources: Centro Atomico Bariloche (CAB), the ENDF/B \cite{endf8} and JEFF \cite{jeff} evaluated nuclear data libraries, or
developed in house. As an example, Fig. \ref{xs_bi} shows total cross section
calculations for bismuth as a polycrystalline material and as an oriented single crystal 
(\texttt{MF=7/MT=4} only) computed using a modified version of NJOY\cite{macfarlane2017njoy}, which
computes the coherent elastic cross sections using NCrystal\cite{cai2020ncrystal}.

\begin{table*}
\caption{Materials included in the library of thermal scattering kernels} \label{t1}
\begin{tabular}{lll}
\hline
Aluminum\textsuperscript{\textasteriskcentered}	&Magnesium Hydride\textsuperscript{\textdagger}	&Polycrystalline Bismuth\textsuperscript{\textdaggerdbl}\\
Iron\textsuperscript{\textasteriskcentered}	&Diamond Nanoparticles\textsuperscript{\textdagger}	&Oriented Single Crystal Bismuth\textsuperscript{\textdaggerdbl}\\
Solid Methane\textsuperscript{\textasteriskcentered}	&Liquid Ethane\textsuperscript{\textdagger}	&Polycrystalline Lead\textsuperscript{\textdaggerdbl}\\
Graphite\textsuperscript{\textasteriskcentered}	&Solid Deuterium\textsuperscript{\textdagger}	&\\
Beryllium\textsuperscript{\textasteriskcentered}	&Liquid Deuterium\textsuperscript{\textdagger,\textasteriskcentered\textasteriskcentered}	&\\
	&Liquid Hydrogen\textsuperscript{\textdagger,\textasteriskcentered\textasteriskcentered}	&\\
	&Light Water\textsuperscript{\textdagger,\textasteriskcentered}	&\\
	&Heavy Water\textsuperscript{\textdagger,\textasteriskcentered}	&\\
\hline
\multicolumn{3}{l}{Source:}\\
\multicolumn{3}{l}{\textsuperscript{\textdagger} CAB, \textsuperscript{\textasteriskcentered} ENDF/B, \textsuperscript{\textasteriskcentered\textasteriskcentered} JEFF, \textsuperscript{\textdaggerdbl} NJOY-NCrystal}\\
\end{tabular}
\end{table*}

With the exception of solid deuterium, all scattering kernels were processed with NJOY2016 in the ACE format with
a continuous outgoing energy distribution (\texttt{iwt=2} ) for use in MCNP 6.2 \cite{werner2017mcnp} and OpenMC \cite{romano2013openmc}. The Monte Carlo code PHITS 3.21 \cite{sato2018features} was also modified for internal use to support the new type of thermal scattering libraries.

\begin{figure}[t]
\resizebox{0.6\textwidth}{!}{\includegraphics{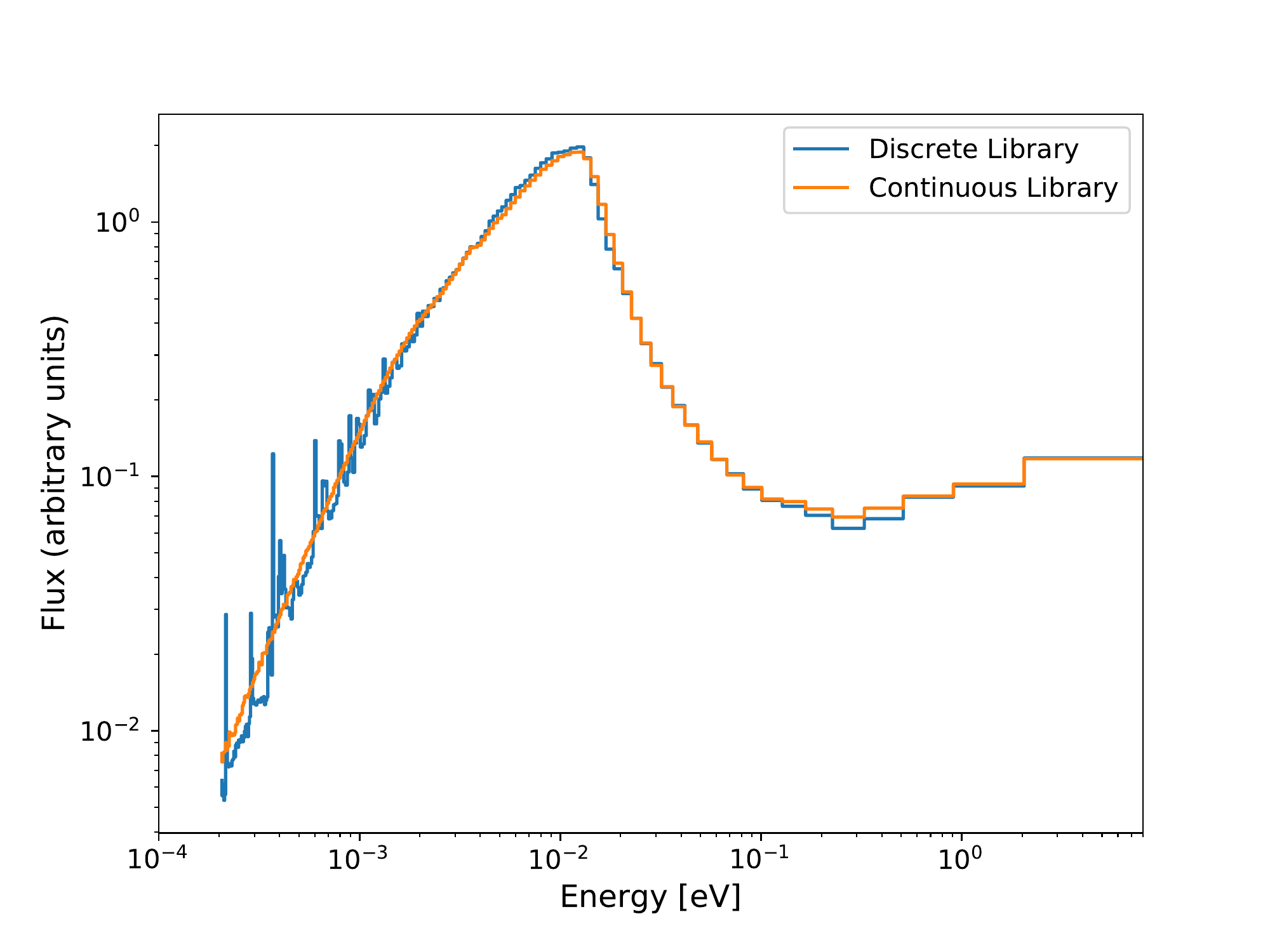}}
\caption{Comparison of spectra from para-hydrogen low dimensional moderator using discrete
and continuous libraries.}\label{discre_cont}
\end{figure}

The use of a continuous distributions for the outgoing energy makes it possible to compute more precisely the
distribution of neutrons in moderators and reflectors \cite{sublet2009accurately}. As an example, Fig. \ref{discre_cont} shows the spectra from a para-hydrogen low dimensional moderator computed using discrete and continuous libraries. The spikes observed at low energies are artifacts caused by the discrete nature of the outgoing spectra \cite{muhrer2004new}.

\begin{figure}[t]
\resizebox{0.6\textwidth}{!}{\includegraphics{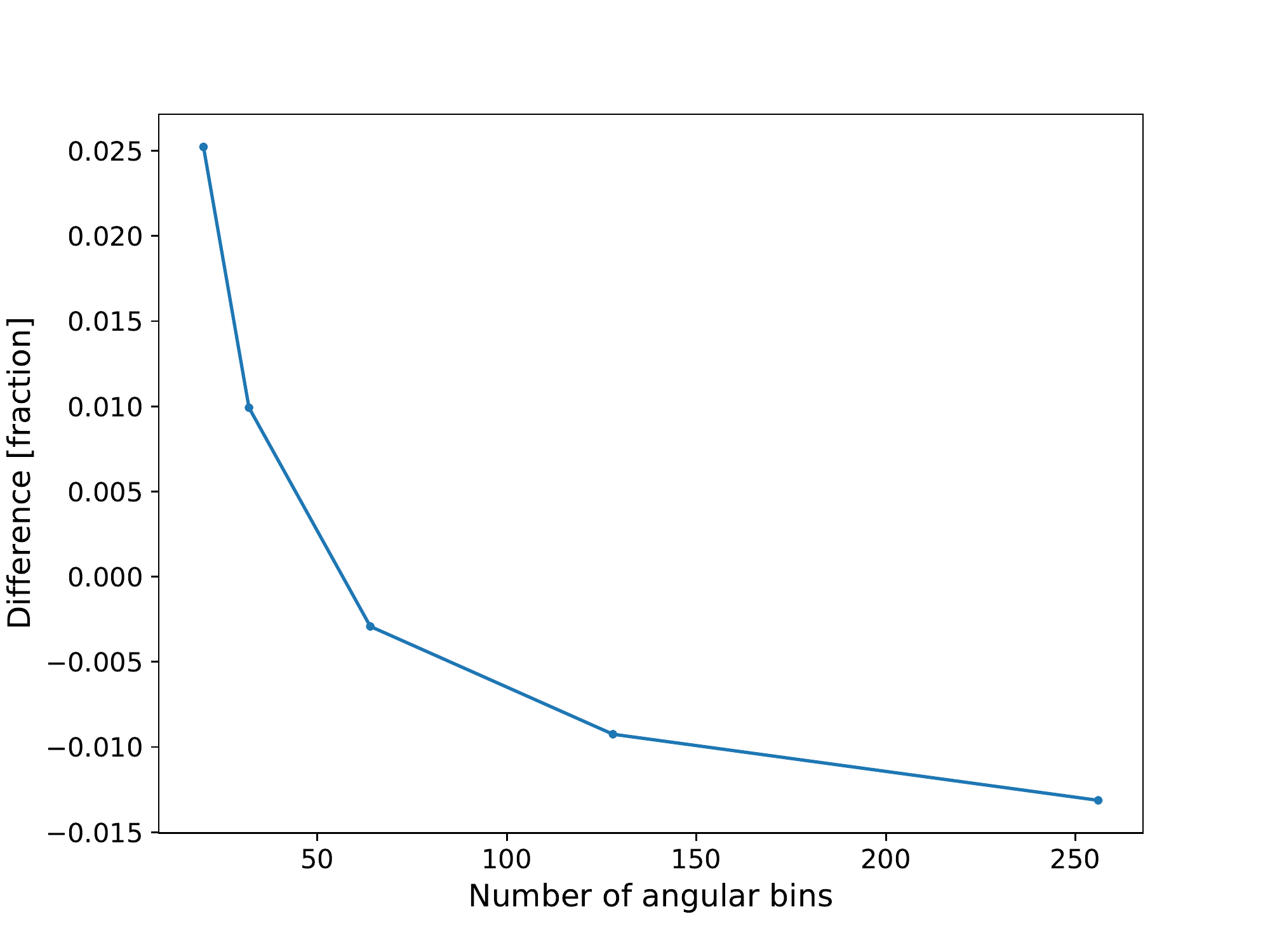}}
\caption{Relative change of the total neutron flux for $E < 20$ meV from a low dimensional moderator, as a function of the number of
angular bins used in the reconstruction of the hydrogen libraries.}\label{diff_conv}
\end{figure}

In addition, the scattering kernels for liquid hydrogen and deuterium were processed with a refined
angular distribution. As shown by Muhrer \cite{muhrer2012urban}, the integral value of the flux emerging 
from a cold neutron source depends on the number of directions used in the processing of the
thermal scattering library. With this in mind, the number of directions selected for the reconstruction 
was 128, achieving a balance between accuracy and size. Overall, changes in integral flux calculations
caused by the use of the new model by Granada \cite{jeff} were compensated in part by the angular effect (Fig. \ref{diff_conv}), resulting in a difference of $\sim1\%$ between the old model, with the coarse angular distribution, and the new model with the fine angular distribution.

\section{Crystallite effect in beryllium}\label{s3}
The impact of crystallite sizes on the neutron scattering properties of a material has been known since the early days of neutron transmission measurements \cite{PhysRev.55.1101,PhysRev.71.589}. 
More recently, such effects were postulated as a contribution to an observed discrepancy between measurements and simulations of a nitrogen-cooled beryllium reflector-filter \cite{muhrer2007}.
This particular effect, referred to here, is known as extinction \cite{sabine1985}, which is directly related to the size of the crystallites in the material. A larger crystallite size will result in a larger portion of the Bragg reflected neutrons being re-scattered back into the direction of the transmitted neutrons, thus lowering the scattering cross-section of the material in the Bragg scattering regime. 

\begin{figure}[t]
\resizebox{0.6\textwidth}{!}{\includegraphics{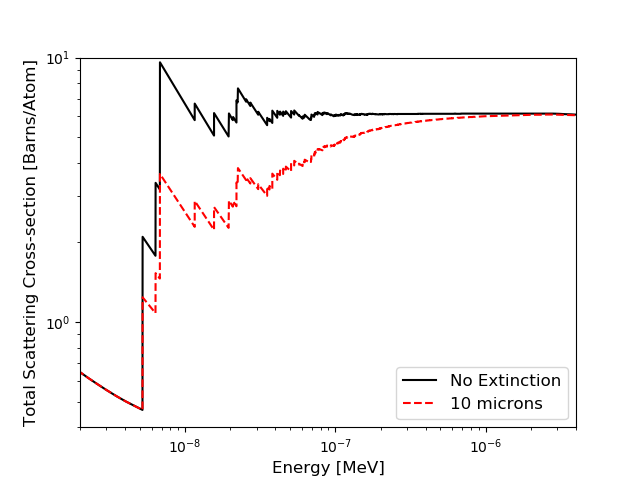}}
\caption{Total neutron scattering cross-section for a beryllium reflector, with and without the extinction effects.}\label{be_ext}
\end{figure}

In order to estimate the impact of such an effect on the cold neutron brightness of the moderator at ESS, we reported earlier on an implementation of an extinction model into the neutron scattering library generation process \cite{dijulio2020}, using the NJOY-NCrystal tool. A selection of the results showing the impact on the total scattering cross-section for beryllium with crystallites of 10 microns in size are shown in Fig. \ref{be_ext}. As described above, the introduction of the finite crystallite size results in a decrease of the total scattering cross-section in the Bragg scattering energy regime. Such an effect was found to result in a lower than 3$\%$ impact on the average cold neutron brightness of the moderators at the facility \cite{dijulio2020}. Measurements on representatives samples of beryllium reflector materials are currently planned and will provide a benchmark for the approach described here. 

\section{New evaluations for hydrogen and deuterium}\label{s4}
A series of recent papers have shown that it is possible to accurately calculate the neutron scattering cross-sections for liquid hydrogen and deuterium based on state-of-the-art molecular modelling techniques \cite{guarini2015,guarini2016}. The use of molecular modelling techniques in the library generation process has become common practice today, however it has not been done for liquid hydrogen and deuterium to this point. 
The simulation of the structure and dynamics for these materials requires the usage of specialized techniques, such as centroid molecular dynamics (CMD) \cite{cao1994a,cao1994b} and ring polymer molecular dynamics (RPMD) \cite{craig2004}, for which software has not been widely available previously. For this reason, we have investigated the usage of the recently released and freely available package, i-PI \cite{iPi2014}, which contains implementations of such molecular modelling techniques, to produce input for the generation of thermal neutron scattering libraries for liquid hydrogen and deuterium. 

\begin{figure}[t]
\resizebox{0.6\textwidth}{!}{\includegraphics{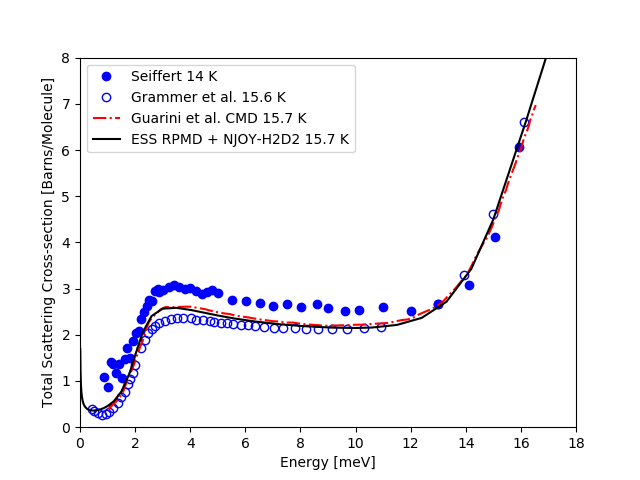}}
\caption{Total neutron scattering cross-section for liquid para-hydrogen from the molecular dynamics techniques described here, Ref. \cite{guarini2015}, and experiments \cite{Seiffert1970,grammer2015}. The data from the previous experiments and simulations was digitized based on the figure presented in \cite{guarini2015} }\label{pH2}
\end{figure}

Our initial results have been presented in \cite{nordin2020}, where more details of the calculation process can be found. Fig. \ref{pH2} shows a slightly revised version of these calculations, compared to previous experimental data \cite{Seiffert1970,grammer2015} and the simulations described in \cite{guarini2015}. To generate our cross-sections, we have combined the molecular dynamics simulations with an implementation of Sk\"old approximation \cite{skold1967} in NJOY, described in \cite{volume42}, and referred to here to as NJOY-H2D2 It can be seen in the figure that our approach can reproduce reasonably well the cross section as computed and presented in \cite{guarini2015}, which demonstrates the possibility of creating neutron scattering libraries for Monte-Carlo simulations, based on such molecular modelling techniques. We are currently working on an implementation for liquid deuterium and exploring the feasibility of using such techniques for other materials, such as hydrogen-deuteride. These will be the topics of an upcoming study \cite{huusko2021}.

\section{HighNESS project - Work Package 2}\label{s5}

In October of 2020, a project to design a high-intensity neutron source at the ESS \cite{santoro2020} was initiated under a European Union Horizon2020 infrastructure design program \cite{highNESS}. The main objective of the project, titled HighNESS, is the design of a second neutron source, placed below the spallation target at the ESS, with the aim of providing a high-intensity of cold, very-cold and ulta-cold neutrons for various scientific endeavours. At the center of the new source is envisioned a large liquid deuterium moderator, which would both provide neutrons directly to beamlines and also to ultra-cold and very-cold neutron sources placed around the moderator location. Additionally, the project will investigate the application of advanced reflector systems to enhance the available neutron fluxes for scientific studies. 

For some of the materials to be investigated, there currently exists no scattering kernels available for use in Monte-Carlo design studies. For this reason, a dedicated work package, encompassing the relevant software development related to the generation of new scattering kernels, was included as a main component of the project. In the initial stages, the work package will re-visit materials for which some work on scattering kernels do exist. These include for example both magnesium hydride \cite{granada2020,muhrer2011} and nanodiamonds \cite{granada2020} for reflector applications. In the latter stages of the project, the work package will investigate the creation of neutron scattering kernels for completely new materials, such as intercalated graphites and clathrate hydrates \cite{zimmer2016}. This workpackage is led by ESS and carried out in collaboration with University of Milano-Bicocca. 

A main focus of the work package is the extension of the freely-available software NCrystal to include new physics models for the proper description of neutron scattering in some of the materials of interest. This for example covers small-angle neutron scattering in nanodiamonds and magnetic scattering in the clathrate hydrates. Additionally, the work package will investigate the integration of the new neutron scattering kernels with Monte-Carlo simulation software. Lastly, input for the new neutron scattering kernels will be generated using state-of-the art molecular modelling techniques, either based on density-functional theory or classical molecular dynamics.
 
 %
 %
 
\section{Modelling of nanodiamonds in Monte Carlo transport}\label{s6}

As a preliminary model for the calculations required in the HighNESS project, we implemented the nanodiamond model from Granada et. al \cite{granada2020} as an internal patch to MCNP and PHITS. In this model, Bragg scattering and inelastic scattering were computed with NJOY-NCrystal, and SANS was modeled with a piecewise power function that approximates the structure factor measured by Teshigawara \cite{teshigawara2019measurement}. 

The scattering kernel is processed with the THERMR and ACER modules of NJOY into an ACE file. The parameters for the SANS model are stored at the end of the ACE file, and a modified version of the Monte Carlo codes read these parameters, computes the SANS and total scattering cross section, and samples the outgoing direction. This modification was implemented in MCNP 6.2 and PHITS 3.21.

\begin{figure}[t]
\resizebox{0.38\textwidth}{!}{\includegraphics{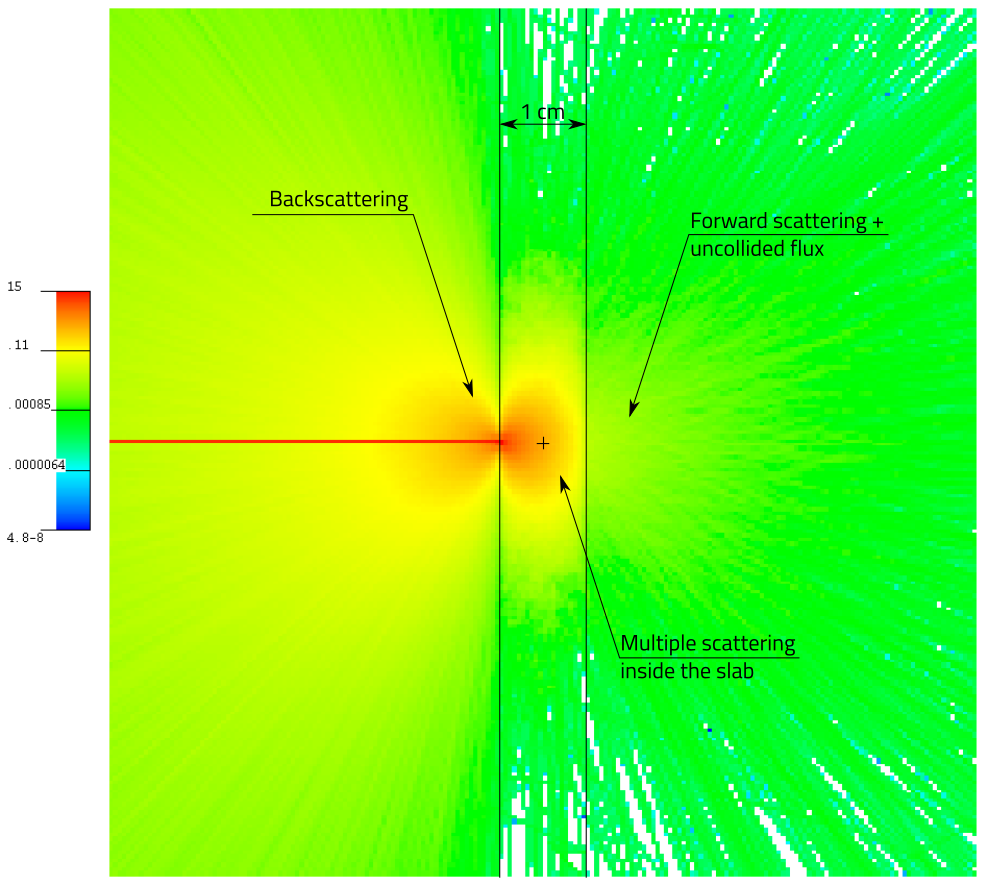}}~~~\resizebox{0.3\textwidth}{!}{\includegraphics{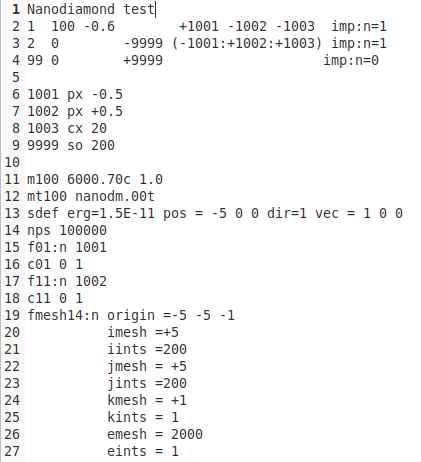}}
\caption{Flux distribution caused by a $0.015$ meV pencil beam coming from the left on a $1.0$ cm slab of nanodiamonds (left). MCNP input file used in this example (right)}\label{nanodm}
\end{figure}

From the user perspective it is only necessary to include the \texttt{nanodm.00t} thermal scattering file, which contains all components for the model (Fig. \ref{nanodm}, right). The results show the increase of the interaction probability caused by SANS, and the Monte Carlo code handles the multiple scattering effect (Fig. \ref{nanodm}, left).

\section{Improved evaluation for hydrogen bound in light water}\label{s7}

The thermal scattering evaluation for light water based on the CAB Model \cite{damian2014cab} is the most advanced evaluation for H$_{2}$O currently available in the nuclear data libraries. In particular, it reduces significantly the differences compared to experimental total cross sections \cite{damian2014cab}, contributes to an elimination of a 500 pcm bias in criticality calculations of plutonium solution benchmarks \cite{brown2018endf}, and helps to explain an anomalous behavior of the temperature evolution of the total cross section \cite{damian2020experimental}.

Nevertheless, the discrete nature of the molecular simulations makes this library a collection of evaluations, instead of a single evaluation. To overcome this, we developed a new evaluation methodology in collaboration with Rolando Granada from Centro At{\'o}mico Bariloche, and Danila Roubtsov from the Canadian Nuclear Laboratories.

\begin{figure}[t]
\resizebox{0.6\textwidth}{!}{\includegraphics{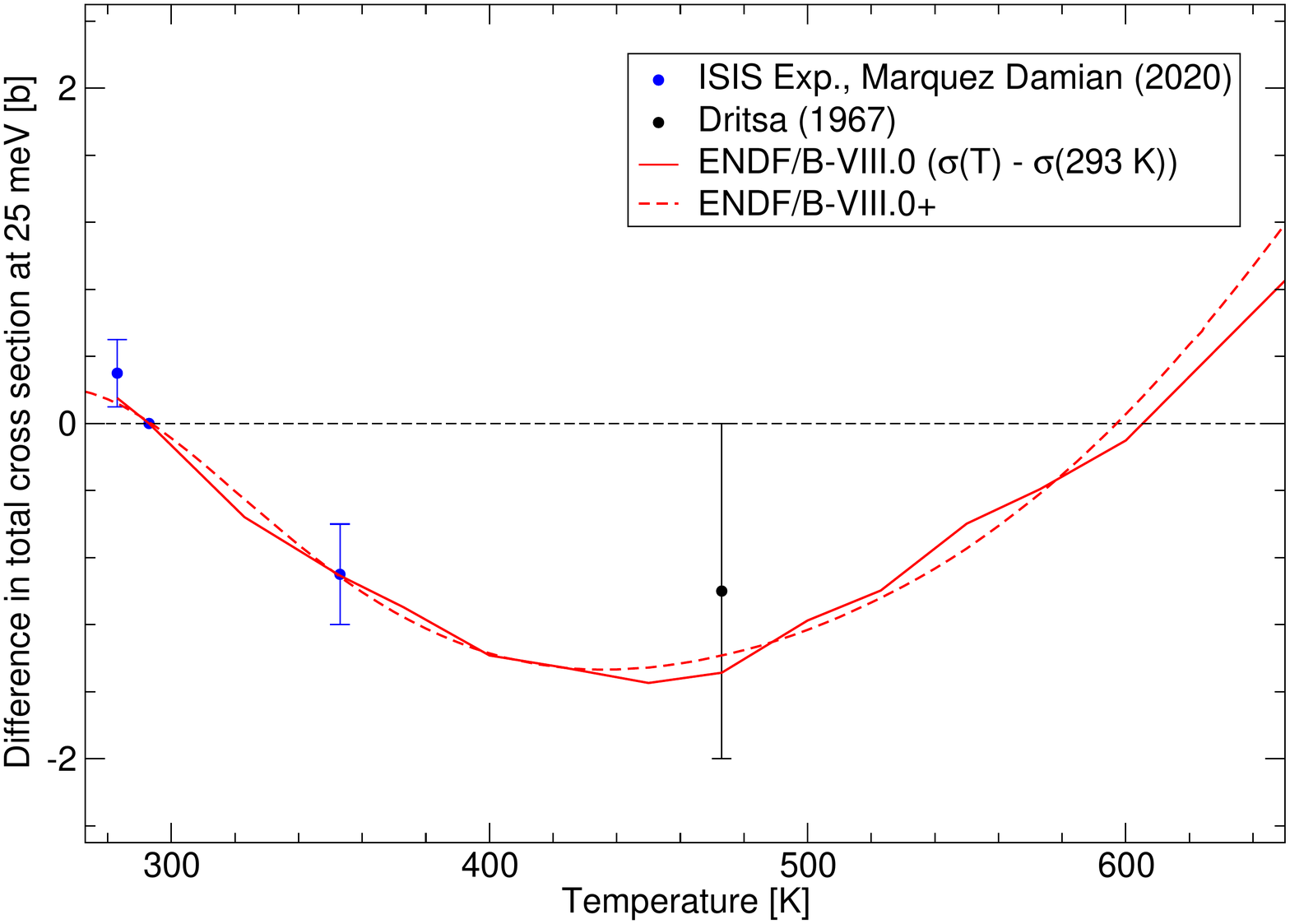}}
\caption{Difference in total cross section of light water at $25$ meV compared to the experimental data from Marquez Damian\cite{damian2020experimental} and Dritsa\cite{dritsa1967}. Calculations with the new model are labelled ENDF/B-VIII.0+}\label{diff_25meV}
\end{figure}

\begin{figure}[t]
\resizebox{0.6\textwidth}{!}{\includegraphics{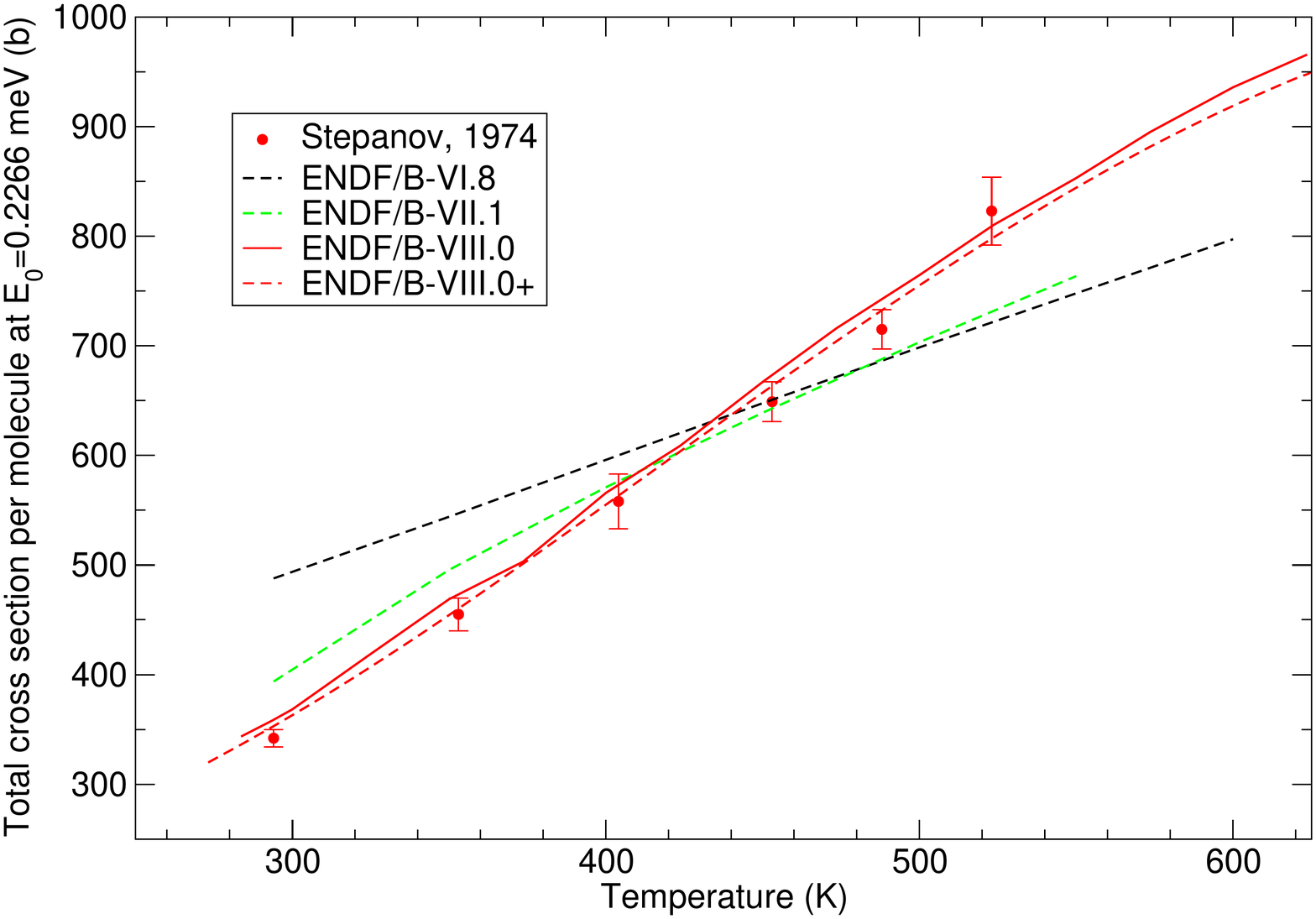}}
\caption{Total cross section at 0.2266 meV computed with different models and compared with measurements by Stepanov \cite{stepanov1, stepanov2}. Calculations with the new model are labelled ENDF/B-VIII.0+}\label{xs_02meV}
\end{figure}

In this new methodology, the computed vibrational spectra for all temperatures is decomposed as a sum of Gaussian distributions, following the work of Esch \cite{esch1971temperature}, Lisichkin \cite{lisichkin2005temperature} and Maul \cite{maul2018perturbation}. The parameters of these Gaussians are fitted with quadratic functions in temperature, under the assumptions that no discontinuities are expected in the liquid phase. These fitted functions make possible to reconstruct the parameters for the evaluation at any temperature. As a result, the scattering cross section has a smoother dependence with temperature and compares well with measurements at thermal (Fig. \ref{diff_25meV}) and cold (Fig. \ref{xs_02meV}) neutron energies. 

In absence of better experimental data, diffusion parameters have recently resurfaced \cite{holmes2021validation} as a possible benchmark for temperature dependence. We compared measured data with computed values, and the new evaluation preserves the good behaviour shown by the CAB Model used in ENDF/B-VIII.0 (Fig. 8), which in turn shows an improvement over ENDF/B-VII and ENDF/B-VI calculations. Nevertheless, such an integral quantity has to be used with care and it is not sufficient to validate a model or the dynamical description that it represents. If the diffusion length is considered alone as a test, one could come to the erroneous conclusion that a simplified model, such as the Synthetic Scattering Function from Granada \cite{granada1986neutron} (dashed black lines in Fig. 8), is better than modern evaluations.

The new evaluation and processed files in ACE format  have been released publicly in a Github repository (\texttt{\href{https://github.com/marquezj/tsl-HinH2O}{https://github.com/marquezj/tsl-HinH2O}}) in a $5$ K temperature grid. The interpolation and reconstruction script, implemented in Python, is also available in the repository.

\begin{figure}[t]
\resizebox{0.6\textwidth}{!}{\includegraphics{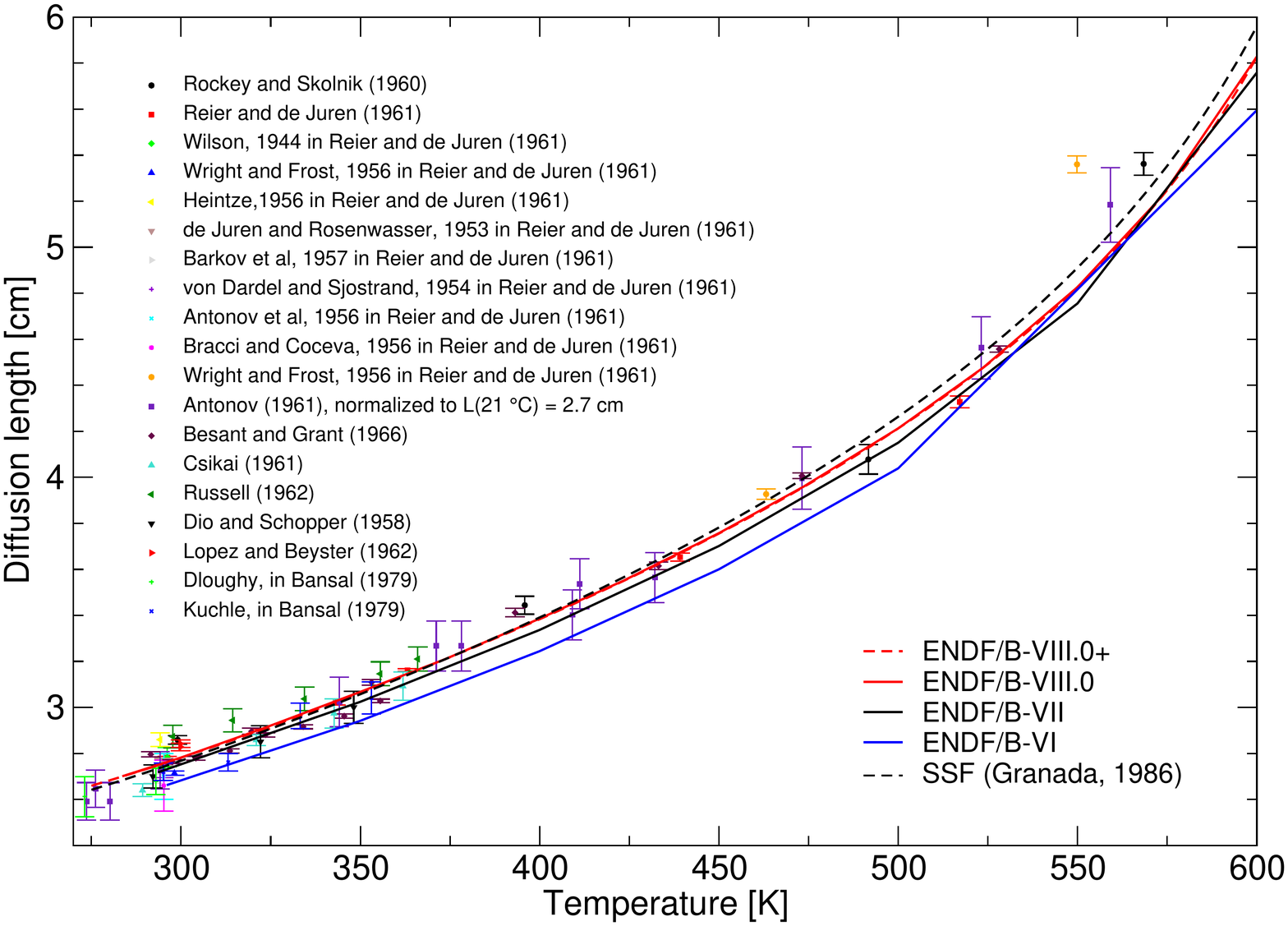}}
\caption{Calculation of the diffusion length as a function of temperature with the new model, compared with experimental data from various
authors \cite{dif1rockey1960measurements,dif2reier1961diffusion,dif3besant1966diffusion,dif4antonov1962investigation,dif5csikai1961measurements,dif6russell1962thermal,diff7dio1958temperature,dif8lopez1962measurement,dif9bansal1974dynamics}, calculations performed with the ENDF/B libraries versions VI, VII, VIII.0 and the Synthetic Scattering Function from Granada\cite{granada1986neutron}. Calculations with the new model are labelled ENDF/B-VIII.0+ (red dashed lines) and are almost equal to the results with ENDF/B-VIII.0 (continuous red line).} \label{diff_length}
\end{figure}

%
%

\section{Acknowledgements}\label{s8}

HighNESS is funded by the European Union Framework Program for Research and Innovation Horizon 2020, under grant agreement 951782.





\nocite{*} 
\bibliographystyle{ios1}           
\bibliography{bibliography}        

%

\end{document}